# Decoding the dark proteome: Deep learning-enabled discovery of druggable enzymes in *Wuchereria bancrofti*

Shawnak Shivakumar[1*], Jefferson Hernandez[2]

[1] Menlo-Atherton High School, Atherton, California
[2] Department of Computer Science, Rice University, Houston, Texas

* Corresponding author
E-mail: shawnak.shivakumar@gmail.com

**Word count:** 3099

**Funding:** The author(s) received no specific funding for this work.

**Conflicts of interest:** All authors declare no conflict of interest.

# Decoding the dark proteome: Deep learning-enabled discovery of druggable enzymes in *Wuchereria bancrofti*

Shawnak Shivakumar[1], Jefferson Hernandez[2]

[1] Menlo-Atherton High School, Atherton, California
[2] Department of Computer Science, Rice University, Houston, Texas

**Abstract**
*Wuchereria bancrofti*, the parasitic roundworm responsible for lymphatic filariasis, permanently disables over 36 million people and places 657 million at risk across 39 countries. A major bottleneck for drug discovery is the lack of functional annotation for more than 90% of the *W. bancrofti* "dark" proteome, leaving many potential targets unidentified. In this work, we present a novel computational pipeline that converts *W. bancrofti's* unannotated amino acid sequence data into precise four-level Enzyme Commission (EC) numbers and drug candidates. We utilized a DEtection TRansformer (DETR) to estimate the probability of enzymatic function, fine-tuned a hierarchical nearest neighbor EC predictor on 4,476 labeled parasite proteins, and applied rejection sampling to retain only four-level EC classifications at 100% confidence. This pipeline assigned precise EC numbers to 14,772 previously uncharacterized proteins and discovered 543 EC classes not previously known in *W. bancrofti*. A qualitative triage emphasizing parasite-specific targets, chemical tractability, biochemical importance, and biological plausibility prioritized six enzymes across five separate strategies: anti-Wolbachia cell-wall inhibition, proteolysis blockade, transmission disruption, purinergic immune interference, and cGMP-signaling destabilization. We curated a 43-compound library from ChEMBL & BindingDB and co-folded across multiple protein conformers with Boltz-2. All six targets exhibited at least moderately strong predicted binding affinities (<1 μM), with moenomycin analogs against peptidoglycan glycosyltransferase and several NTPase inhibitors showing promising nanomolar hits and well-defined binding pockets. While experimental validation remains essential, our results provide the first large-scale functional map of the *W. bancrofti* dark proteome and accelerate early-stage drug development for the species and related parasites.

# 1. Introduction

*Wuchereria bancrofti*, the parasitic roundworm responsible for approximately 90% of lymphatic filariasis (LF) cases, causes one of the most debilitating and widespread neglected tropical diseases. An estimated 51 million people are infected globally, with 36 million suffering permanent disfigurement that severely impairs mobility and quality of life. As of 2023, an additional 657 million individuals across 39 countries are at risk, making LF a major contributor to long-term disability worldwide [1,2]. Despite global elimination efforts, current standard of care drugs such as diethylcarbamazine or ivermectin only effectively clear larval worms but do not reliably kill adult worms, requiring repeated treatments. With no vaccine and limited drug efficacy, new treatments targeting adult worms are urgently needed [3].

A significant obstacle to developing new treatments lies at the molecular level: the parasite's proteome is poorly understood. The recently sequenced *W. bancrofti* genome is predominantly "dark", with 10% of nearly 11,000 genes functionally annotated [4]. Most proteins in the Universal Protein Resource (UniProt) are labeled "uncharacterized," highlighting a considerable gap in functional annotation [5]. Without knowing the function of 90% of the roundworm's proteins, identifying new therapeutic targets remains extremely challenging.

Closing this annotation gap is both a scientific and a humanitarian imperative. The emergence of deep learning offers a powerful opportunity: modern protein function prediction models can decode patterns in amino acid sequences that hint at enzymatic activities or biochemical roles, even in the absence of experimental data. The Enzyme Commission (EC) system classifies enzymes by reactions using four hierarchical levels (e.g., EC 3.4.22.1), where the first digit indicates reaction type and subsequent digits specify substrate and mechanism [6].

Assigning EC numbers yields specific information about catalytic roles and druggable sites in pathogens. Unlike vague annotations, EC numbers define direct biochemical activities, making them invaluable for predicting enzymatic functions [7].

The main objective of our study is to use deep learning to prioritize druggable targets and identify small-molecule inhibitors against *W. bancrofti*. We report a substantial expansion of the *W. bancrofti* enzyme repertoire, uncovering hundreds of previously hidden enzymes and assigning them putative functions. From these, we highlight six priority targets spanning diverse biology—from bacterial symbiosis to host immune evasion—each of which can be inhibited by small molecules in the low-micromolar to nanomolar range. In summary, we demonstrate a rapid framework from unannotated pathogen proteomes to drug target candidates, which is especially valuable for neglected tropical diseases.

## 2. Methods

Recent deep learning advances have greatly improved EC prediction accuracy over classical homology-based methods [8]. We leveraged these advances to systematically annotate the *W. bancrofti* proteome using an end-to-end pipeline visualized in Figure 1 that first distinguishes enzymes from non-enzymes using a modified binary sequence classifier, then assigns EC numbers to enzymes using the Hierarchically-Finetuned Nearest Neighbor (HiFi-NN) model [9]. After annotating thousands of parasite enzymes, we devised a multi-criteria prioritization strategy and focused on a handful of high-value enzymatic targets spanning various mechanistic classes. Finally, to evaluate the druggability of these targets, we performed structure-based virtual screening using Boltz-2: a state-of-the-art AI-based molecular docking (co-folding) model [10].

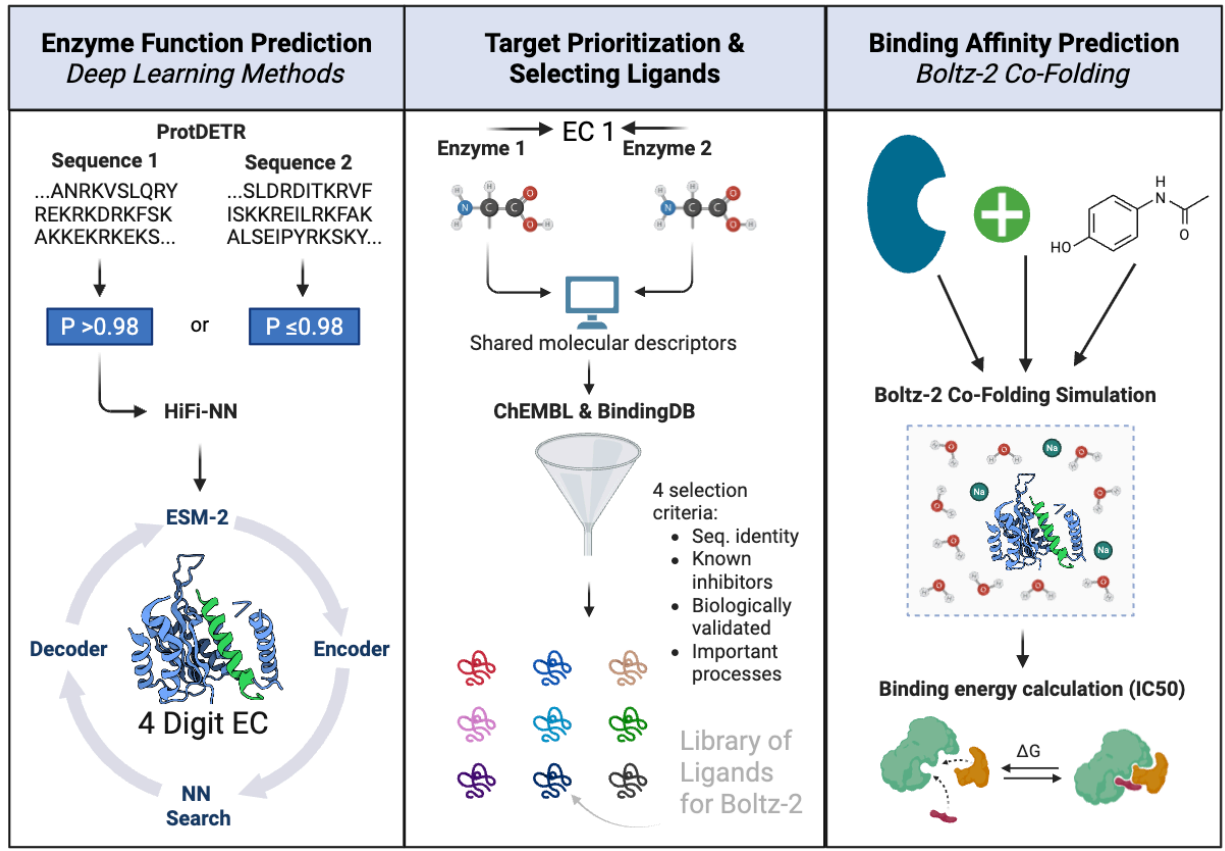

**Figure 1.** Schematic overview of the computational pipeline used for enzyme annotation, target prioritization, and inhibitor screening in *W. bancrofti*. ProtDETR [11], HiFi-NN / ESM-2 [9] predict four-digit EC numbers, followed by target selection using ChEMBL [14] and BindingDB [15] based on sequence identity, known inhibitors, validation status, and biological importance. Boltz-2 [10] co-folding simulations then predict binding poses and IC$_{50}$ values to identify promising candidate pairs.

The pipeline shown in Figure 1 allows us to test a curated set of candidate inhibitors against each target and predict binding affinities and poses. In doing so, we aimed to identify novel parasite enzymes and specific protein–inhibitor pairs as potential starting points for anti-filarial drug development.

*2.1 Data Acquisition and Preprocessing*

We obtained the *W. bancrofti* proteome from the UniProt Database (Taxonomy ID: 6293) [5]. The amino acid sequence of each entry was stored in FASTA format with its UniProt accession. Data curation involved the removal of duplicate sequences and those containing non-standard amino-acid characters.

After filtering, we were left with 47,788 *W. bancrofti* proteins, of which 4,476 had EC annotations. We designated these 4,476 labeled sequences as the training set used to fine-tune the last layer of our EC predictor and for calibration of decision thresholds. The remaining 43,312 unannotated sequences served as the input to the annotation pipeline.

*2.2 Enzyme vs Non-Enzyme Binary Classification*

The first step of the pipeline classified each protein in the *W. bancrofti* proteome as either an enzyme or a non-enzyme. We used Protein Detection Transformer (ProtDETR) [11] as the starting architecture since it natively performs residue-level detection of functional segments through a bank of learned queries that highlight catalytic patterns and output EC-oriented scores with localization maps. As shown in Figure 2, we adapted this capability to produce one confidence probability per sequence by using a compact evidence-aggregation head that converts ProtDETR's multiple proposals into a single probability of enzymatic function denoted as P(enzyme). The adapter pools the strongest proposals with attention-weighted gating and keeps the underlying localization maps, yielding a high-precision, interpretable gate suitable for proteome-scale screening.

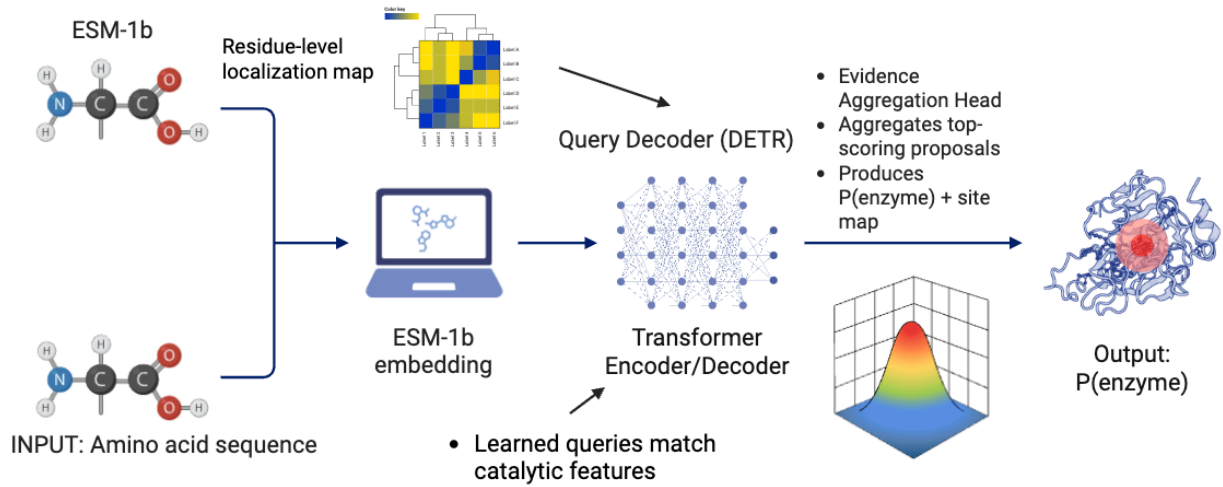

**Figure 2.** Modified ProtDETR Architecture for probability *P(enzyme)* output

The final decision threshold was a strict cutoff of P(enzyme) > 0.98, chosen to have a high confidence of true positives passed to downstream EC assignment.

*2.3 Four-Digit EC Number Assignment with HiFi-NN*

After filtering out non-enzymes, we assigned four-digit EC numbers to the previously uncharacterized proteins. To enable these predictions at the proteome scale, we adapted the HiFi-NN framework, a deep learning architecture that combines embedding from a transformer with a hierarchical nearest-neighbor classifier to infer EC labels with high precision [9]. We fine-tuned the last layer of the HiFi-NN model using the 4,476 *W. bancrofti* proteins with known EC annotations. Our training pipeline included data cleaning and class-rebalancing routines to correct for the extreme skew across EC classes. We trained for 10 epochs with AdamW (learning rate = 1e-4, betas = 0.9/0.999, eps = 1e-8, weight decay = 0.1) at batch size 32 and 32-bit precision; the best checkpoint was chosen by validation loss, and convergence was visualized in Weights & Biases (wandb) [12].

**Table 1.** Performance of HiFi-NN before and after domain-specific fine-tuning on *W. bancrofti*

| Model | Accuracy | Precision | Recall | F1 | AUC | MRR |
|---|---|---|---|---|---|---|
| Baseline | 0.8710 | 0.7328 | 0.7065 | 0.7131 | 0.7137 | 0.8983 |
| **Epoch 4** | **0.8774** | **0.7364** | **0.7225** | **0.7239** | **0.7555** | **0.9026** |
| Epoch 9 | 0.8703 | 0.7172 | 0.7064 | 0.7042 | 0.7443 | 0.8991 |

While top-1 accuracy increased modestly, the Epoch 4 fine-tuned model produced the best combination of area under the curve (AUC) and mean reciprocal rank (MRR), as shown in Table 1, both of which reflect improved rank-ordering and confidence separation among hard-to-distinguish classes. As shown in Figure 3, later epochs exhibited rising validation loss relative to training loss, which is consistent with overfitting. Therefore, the Epoch 4 checkpoint was selected as the backbone for all downstream annotations.

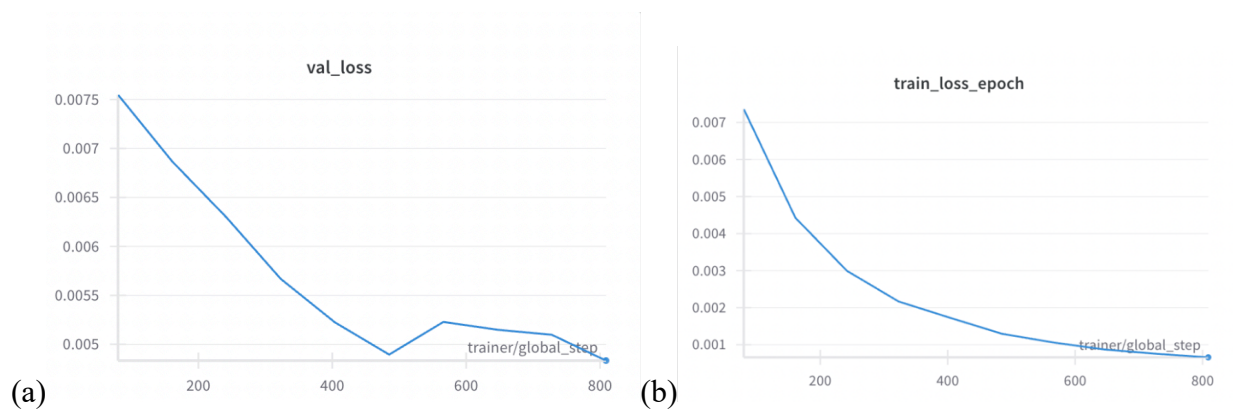

**Figure 3.** Validation loss (a) and training loss (b) time series across fine-tuning epochs in wandb

We then deployed this model to annotate all the sequences that passed the ProtDETR binary enzyme classifier. Each sequence was submitted to the HiFi-NN classifier, and our model returned a ranked list of candidate EC numbers along with associated confidence scores. To maximize reliability, we apply rejection sampling: only predictions with 100% model confidence across all four EC digits were selected for further analysis. We also counted the number of paralogs per EC, which is defined as the count of distinct proteins mapped to that class. This threshold is far stricter than conventional practice but reflects our design goal of prioritizing specificity over recall in early-stage drug target discovery.

*2.4 Target Prioritization Strategy*

Following EC assignment, we stratified enzyme classes by number of paralogs to reflect further annotation confidence and potential biological importance. Each EC was placed into one of three abundance tiers based on the number of paralogs: ≤3, 4–9, and ≥10. For target discovery, we restricted consideration to EC classes with at least 10 independently annotated proteins. This threshold lowers the risk of misannotations and raises confidence in the underlying labels used for target prioritization.

Within this high-confidence subset, we conducted a qualitative, criteria-based review. Candidates advanced only if they satisfied all four methodological checks below:

1. **Parasite-specific target**: We computed BLASTp [13] alignments of the parasite protein against the reviewed human proteome and deprioritized any enzyme with a human homolog exceeding 30% sequence identity. Close human homologs raise the risk of on-target toxicity and drastically narrow the available therapeutics.
2. **Chemically tractable**: We queried ChEMBL [14] and BindingDB [15] to find compounds or inhibitors against the enzyme class in parasites.

3. **Biochemical importance**: The reaction the enzyme catalyzes must play an important role in the parasite, so that inhibition will likely impair survival.
4. **Biologically plausible**: Enzymes with no precedent in similar organisms were excluded; we required that each candidate enzyme be present in some related parasite.

*2.5 Ligand Co-Folding with Boltz-2*

After prioritizing targets, we compiled a list of small-molecule inhibitors for each enzyme class to employ Boltz-2, a co-folding model that predicts protein-ligand complex structures and binding affinities. Boltz-2 is roughly 1000x faster than physics-based methods with similar accuracy, enabling us to screen dozens of compounds quickly [10]. For each complex as visualized in Figure 4, we calculated the predicted affinity value with Boltz-2 and visually inspected the binding pose to remove any structural artifacts. We prioritized targets with high affinity scores and identified specific ligand chemotypes that could serve as initial hits.

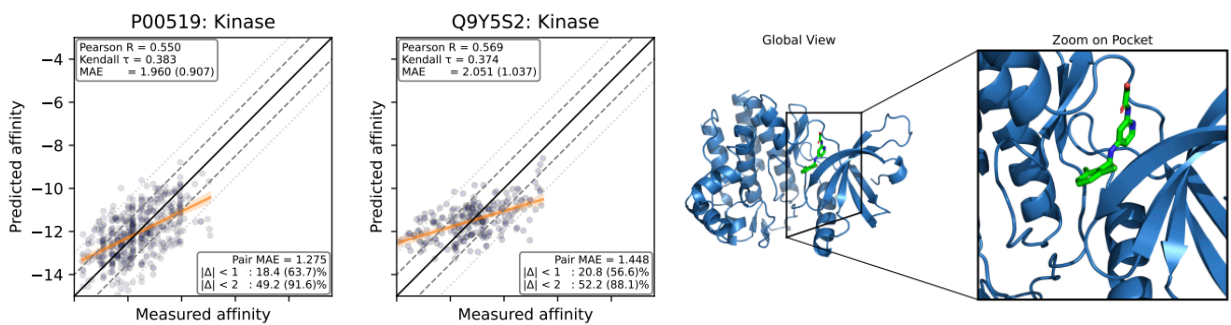

**Figure 4.** Boltz-2 affinity calculation for representative kinases and binding pose visualization

$$\Delta G \approx (6 - \log_{10}(IC_{50})) \times 1.364 \qquad (1)$$

Boltz-2 outputs binding affinity in $\log(IC_{50}\,(\mu M))$, which can be converted to $\Delta G$ (kcal/mol) using equation 1 above [10]. For consistency, we report all results in $IC_{50}$ units.

## 3. Results

*3.1 Proteome-Wide Annotation Outcomes*

Our pipeline dramatically increased functional information for *W. bancrofti,* identifying novel drug targets. Starting from **43,312** unannotated sequences, we obtained the following:

1. **Binary enzyme classification:** 37,171 sequences (~85.8%) were predicted as enzymes with >98% confidence, suggesting an enzyme-rich proteome. This partitioning is informative: *W. bancrofti* appears to have an unusually enzyme-rich proteome, though we acknowledge that some over-prediction may occur due to divergent sequences.
2. **EC number assignments:** We assigned full four-digit EC numbers to **14,772 proteins** using the 100%-confidence filter. Their distribution across the seven EC superclasses is shown in Figure 5; transferases (49%) and hydrolases (37%) dominate, consistent with parasite reliance on molecular modification and degradation.
3. **Novel enzyme classes in *W. bancrofti*:** We identified **543 unique EC numbers** among the predictions not previously associated with *W. bancrofti* in any database or publication. Many of these are typical metabolic enzymes shared with other organisms. Importantly, **73 of these new EC classes have ≥10 distinct UniProt sequences** in the *W. bancrofti* proteome.
4. **Presence of bacteria-like enzymes:** The presence of bacterial enzymes (e.g., the peptidoglycan glycosyltransferase EC 2.4.99.28) was confirmed for sequences that cluster phylogenetically with bacterial homologs. This reflects the symbiotic relationship between *W. bancrofti* and its obligate intracellular bacterium Wolbachia, which supplies essential metabolites, such as heme and riboflavin, that the roundworm cannot produce independently. Eliminating Wolbachia by antibiotics disrupts this symbiotic relationship, leading to sterility and death of *W. bancrofti* [16].

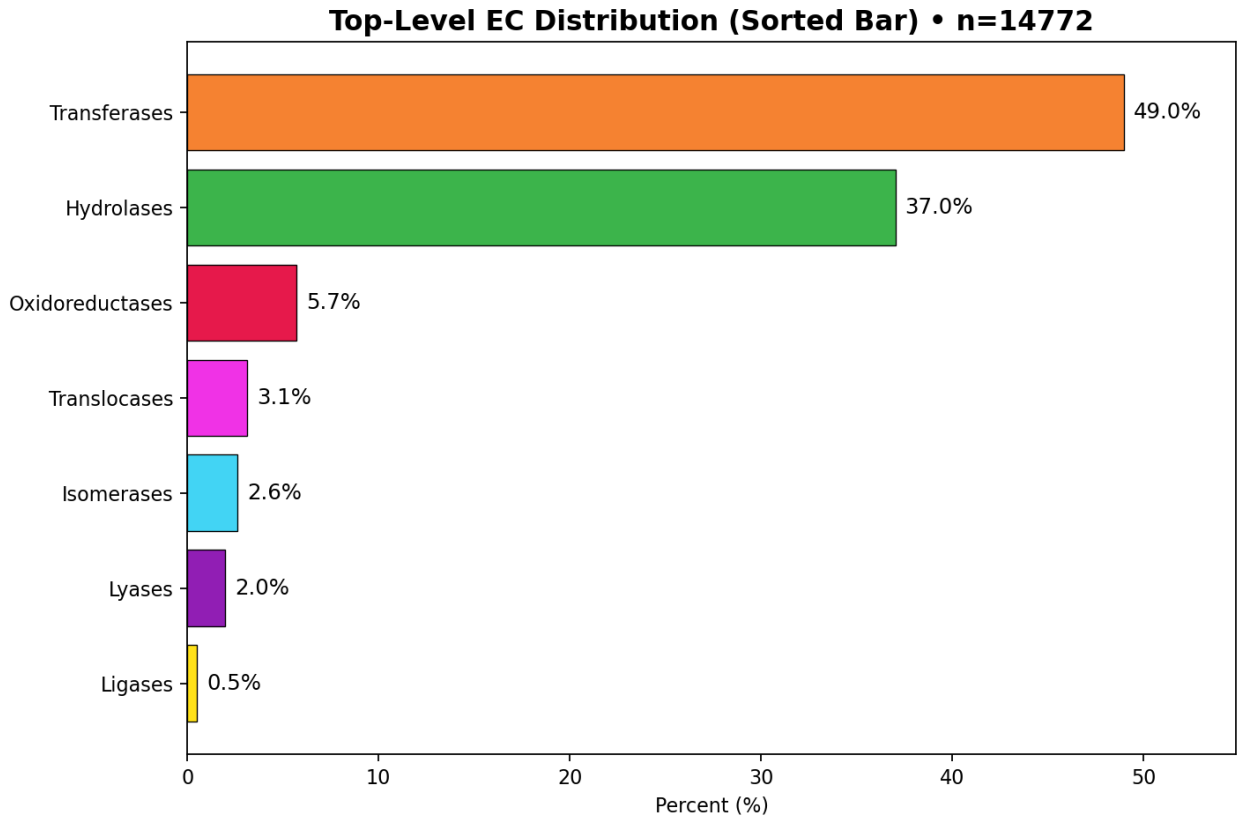

**Figure 5.** Distribution of top-level Enzyme Commission (EC) classes for 14,772 high-confidence enzyme annotations in Wuchereria bancrofti. Categories are sorted by abundance, with transferases (49.0%) and hydrolases (37.0%) dominating the proteome, followed by oxidoreductases (5.7%), translocases (3.1%), isomerases (2.6%), lyases (2.0%), and ligases (0.5%). Percentages indicate the proportion of annotated proteins assigned to each EC superclass. The predominance of catalytic functions involved in group transfer and hydrolytic cleavage is highlighted in this bar plot.

*3.2 Target Prioritization, Selection, and Co-Folding*

We applied our four prioritization filters (Section 2.5) to triage our 73 new EC classes with over 10 paralogs. This process highlighted 5 major strategic categories of drug targets, each represented by one or two top candidate enzymes from our annotations. For predicted binding affinity, we used a standard early-discovery benchmark: IC$_{50}$ ≤ 1 micromolar (µM) qualifies as a hit suitable for hit-to-lead work.

*3.2a Strategy 1 - Anti-Wolbachia*

**Enzyme:** *Peptidoglycan glycosyltransferase* (EC 2.4.99.28) [17]

**Overall Druggability Result:** Strong Hit**.** Assigned to **41** *W. bancrofti* protein sequences.

**Criterion 1, Parasite protein specific:** Global similarity to human homologs is low at **21.80%** sequence identity, which is below our threshold of <30%.

**Criterion 2, Chemical tractability**: Four inhibitor classes have precedent against bacterial glycosyltransferases: Moenomycin A, Neryl-moenomycin analogs, Teixobactin (undecapeptide), and Dimethylaminostyryl-pyranyltropylium [14, 15].

**Criterion 3, Biochemical importance:** Peptidoglycan glycosyltransferase catalyzes the polymerization step in peptidoglycan synthesis – a process essential for *Wolbachia* bacteria [18].

**Criterion 4, Biologically plausible:** Prior studies have shown that anti-Wolbachia therapy can sterilize or kill other adult filarial roundworms over time [16].

As shown in Figure 6, Moenomycin A12 is consistently the top performer (IC$_{50}$ ≈ 16 nM), and Neryl-moenomycin is a close second with low nanomolar affinity across sequences.

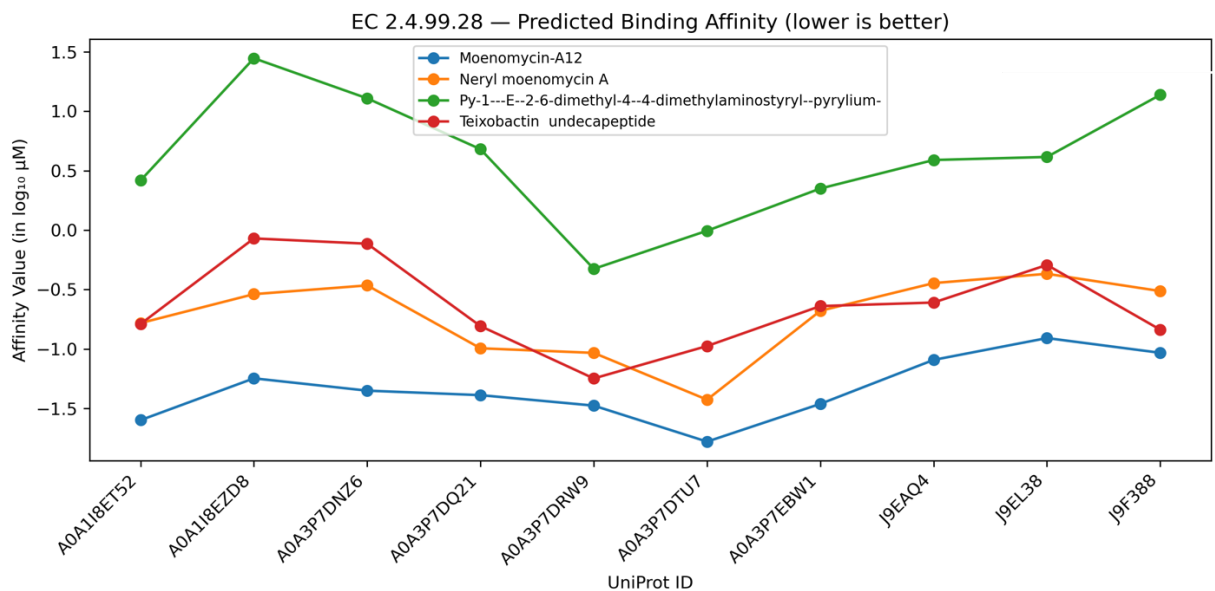

**Figure 6.** Predicted potency of glycosyltransferase inhibitors across 10 *W. bancrofti* sequences

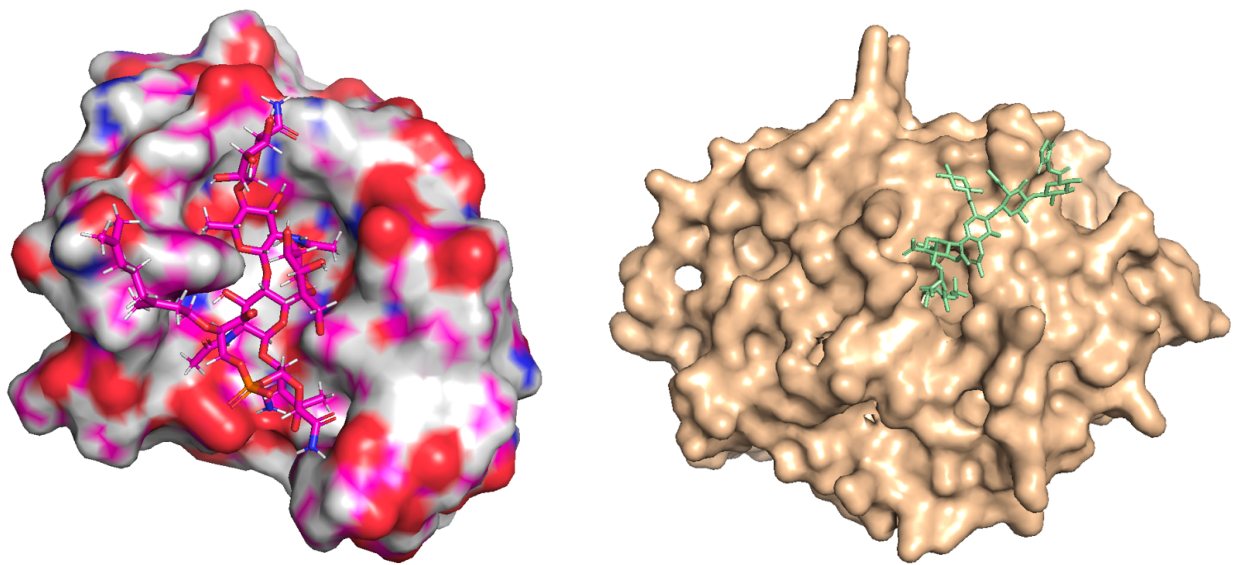

**Figure 7.** Neryl-moenomycin (left) and Moenomycin (right) complexed with glycosyltransferase

The Boltz-2 pose above in Figure 7 shows the ligand spanning the glycosyltransferase groove, hydrogen-bonding to the catalytic residues, and anchoring along the donor/acceptor subsites; the pocket is deep and well-defined, consistent with the strong predicted affinity.

*3.2b Strategy 2 – Targeting core parasite enzymatic reactions*

**Enzyme:** *Cathepsin B* (EC 3.4.22.1) [19] & *Peptidase 1* (EC 3.4.22.65) [20]

**Overall druggability result:** Moderate-strong hit. Assigned across **28** *W. bancrofti* proteins.

**Criterion 1, Parasite-specific target:** Global similarity to human homologs is low at **13.36%** sequence identity, comfortably below our selectivity thresholds of <30%.

**Criterion 2, Chemical tractability**: We assembled a ten-compound reference set spanning covalent and reversible chemotypes commonly used against cathepsins: K777**,** E-64**,** E-64d (aloxistatin)**,** Z-FY-CHO**,** CA-074 Me**,** Nitroxoline**,** VBY-825**,** CLIK-148**,** Relacatib (SB-462795), and Dipeptide nitriles/aldehydes [14,15].

**Criterion 3, Biochemical importance:** These clan CA cysteine endopeptidases likely drive essential parasite processes, including feeding (hemoglobin and tissue protein digestion), cuticle remodeling during molts, and embryogenesis. Inhibiting them disables core machinery that the nematode cannot readily bypass [21].

**Criterion 4, Biologically plausible:** These enzymes are well known in filarial biology – for instance, the *Brugia malayi* nematode has cathepsin B, which is important for development [21].

Across both cysteine-protease families, as shown in Figure 8, K777 is the top performer, with best predictions around (IC$_{50}$ ≈ 250 nM) and frequent sub-micromolar values. VBY-825 and Relacatib show sub-micromolar activity on subsets, whereas E-64/E-64d are largely micromolar and, CA-074 Me is weak in this set.

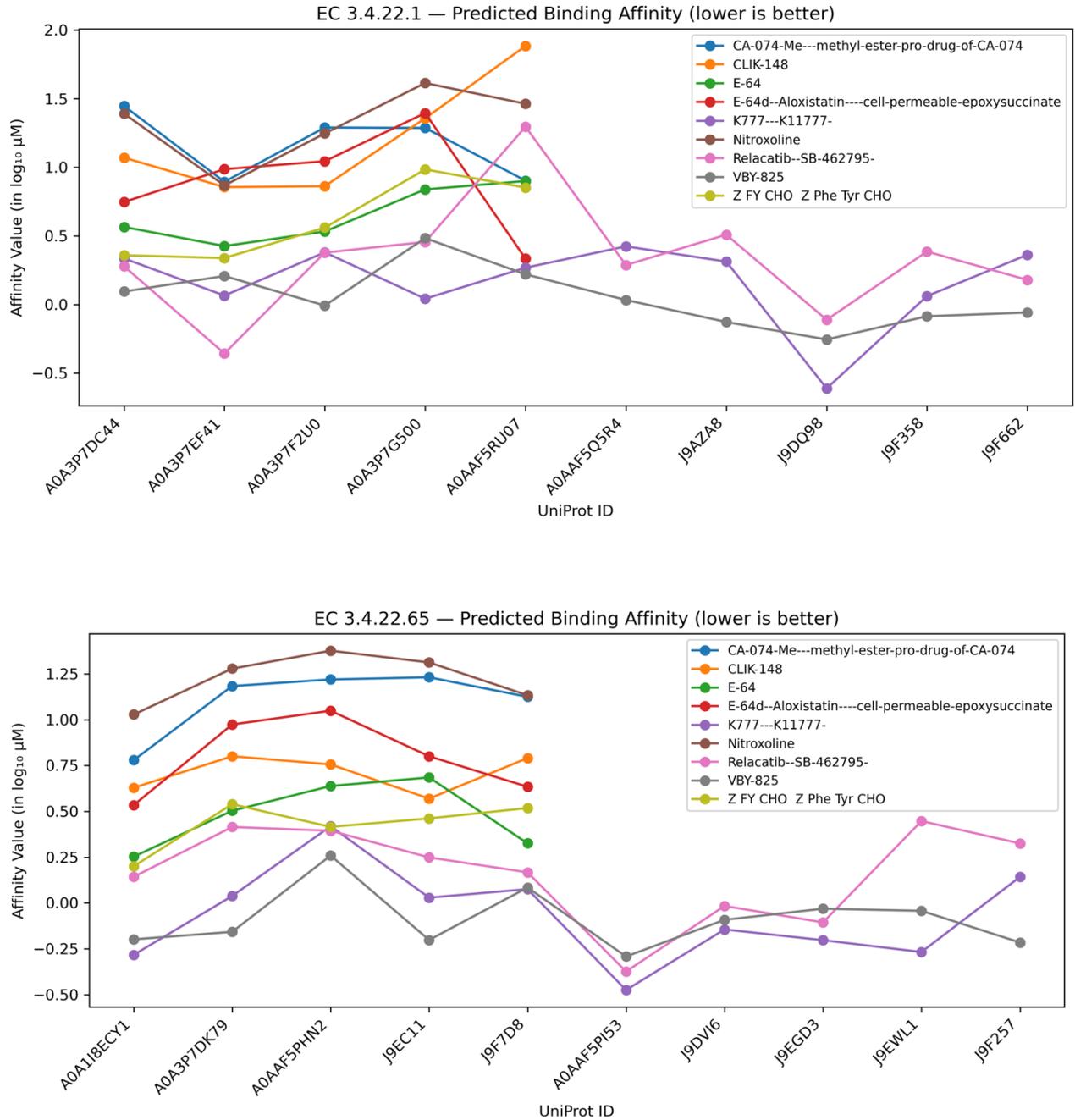

**Figure 8.** Predicted potency of cysteine-protease inhibitors against *W. bancrofti* cathepsins

*3.2c Strategy 3 - Life-Cycle/Transmission Blockers*

**Enzyme:** *Chitinase* (EC 3.2.1.14) [22]

**Overall druggability result:** Moderate hit. Assigned across **49** *W. bancrofti* sequences.

**Criterion 1, Parasite-specific target:** Global similarity to human homologs is low at **10.06%** sequence identity, comfortably below our selectivity thresholds of <30%.

**Criterion 2, Chemical tractability**: We evaluated six inhibitor classes with prior activity against GH18 chitinases: Allosamidin, Argifin, Argadin, Bisdionin C, Bisdionin F, and CI-4 [c-(L-Arg-D-Pro)] [14,15].

**Criterion 3, Biochemical importance:** The GH18 chitinase is likely secreted by microfilariae. Blocking this reaction arrests the life cycle in the vector rather than killing adult roundworms in the human host. As a result, chitinase is best positioned as a population-level transmission-blocking target that complements macrofilaricidal drugs [23].

**Criterion 4, Biologically plausible:** Experimental work in two filarial nematodes, *Brugia* and *Onchocerca,* has shown that blocking or knocking down the larval chitinase prevents the parasite from shedding its sheath, thus blocking its transmission by mosquitoes [24].

As shown in Figure 9, Allosamidin is the best performer with one sub-micromolar prediction and several ~1–5 μM estimates. Overall, the chitinase pocket is chemically addressable but yields moderate predictions, with isolated sub-micromolar values for Allosamidin.

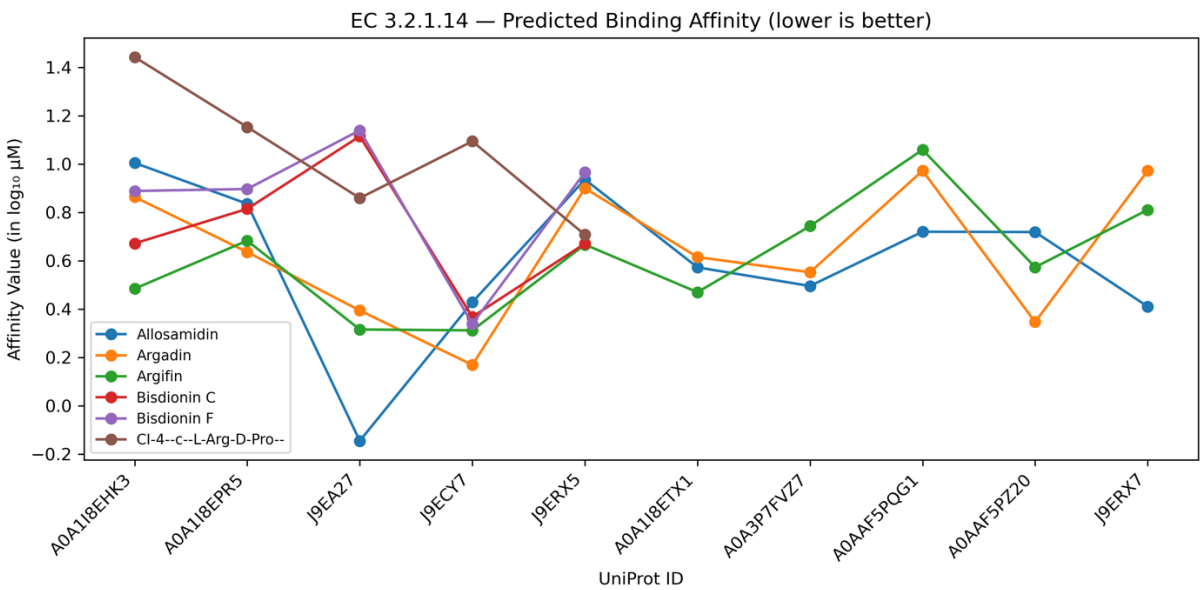

**Figure 9.** Predicted potency of chitinase inhibitors against *W. bancrofti* EC 3.2.1.14

*3.2d Strategy 4 - Immune-Evasion Disruptors*

**Enzyme:** *Nucleoside-triphosphate phosphatase (NTPase)* (EC 3.6.1.15) [25]

**Overall druggability result:** Strong hit. Assigned across **10** *W. bancrofti* sequences.

**Criterion 1, Parasite-specific target:** Global similarity to the closest human homolog is **29.66%** sequence identity, just below our <30% threshold. Similarity of NTPase-conserved regions is expected, but several pocket-lining residues diverge, which offers opportunities for parasite-specific drug design.

**Criterion 2, Chemical tractability**: We screened eight chemotypes with precedent against NTPase: ARL-67156, Reactive Blue 2, Suramin, NF279, PPADS, PSB-06126, PSB-069, CD39-IN-1 (Compound 338) [14,15].

**Criterion 3, Biochemical importance:** This enzyme likely helps the roundworm by degrading ATP released from damaged tissues or platelets, thereby reducing inflammation and preventing blood clotting in lymphatic vessels. Targeting the parasite's NTPase could remove this protective effect, exposing the roundworm to the host immune system and making it vulnerable to other interventions [26].

**Criterion 4, Biologically plausible:** NTPases contribute to virulence and immune modulation in multiple helminths and protozoa [27].

Across three *W. bancrofti* proteins, Reactive Blue 2 and suramin achieve the best results as represented by Figure 10 (≈0.1–0.15 µM), ARL-67156 is generally sub- to low-micromolar, and CD39-IN-1 is weaker and variable.

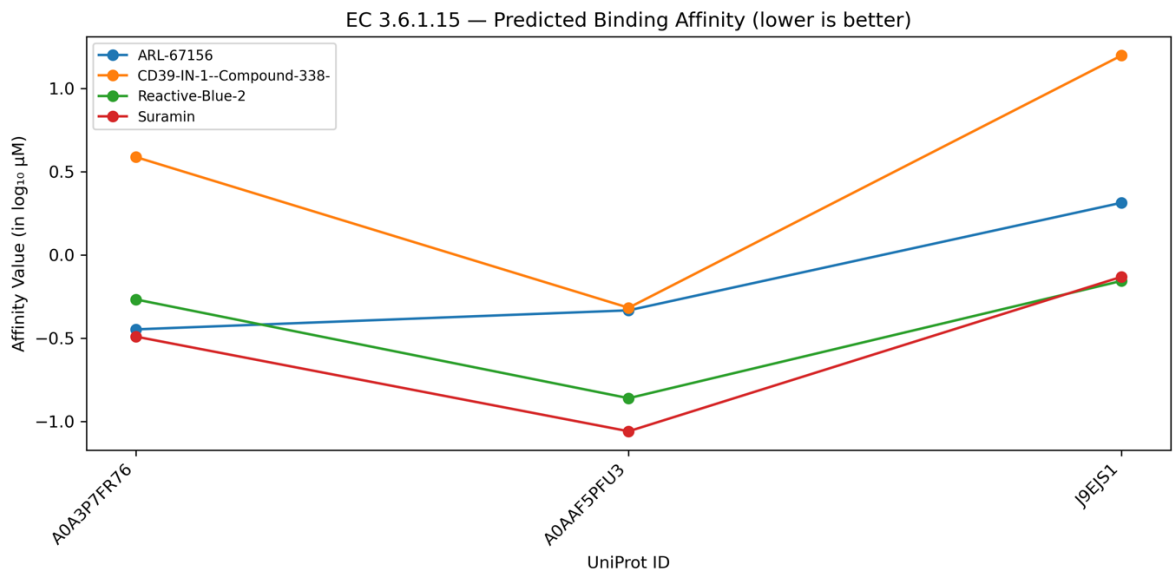

**Figure 10.** Predicted potency of NTPase (EC 3.6.1.15) inhibitors

Overall, the combination of surface accessibility shown in Figure 11, host-selectivity margin, and multiple sub-micromolar starting points supports NTPase as a high-value immune-evasion target.

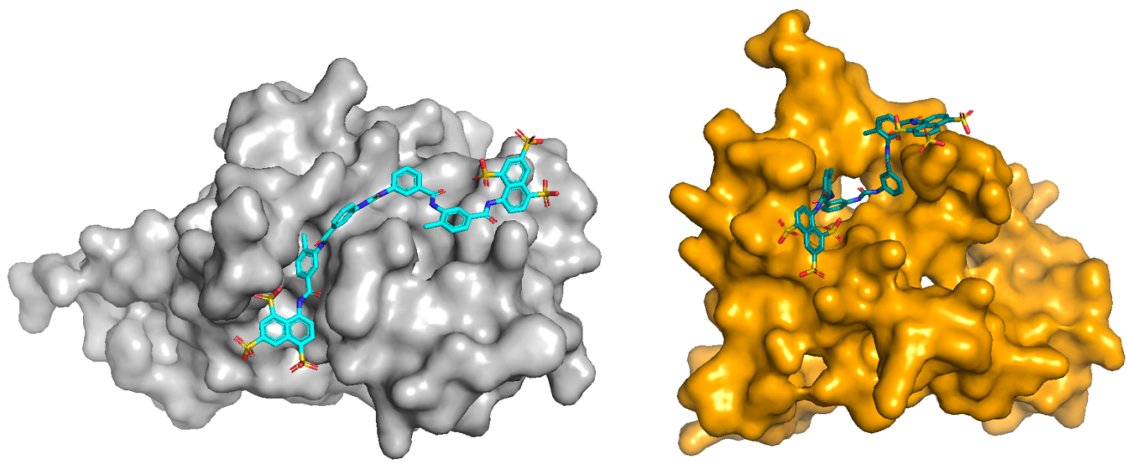

**Figure 11.** Suramin bound to NTPase (EC 3.6.1.15)

*3.2e Strategy 5 - Metabolic/neuromuscular destabilization*

**Enzyme**: *3',5'-Cyclic-GMP Phosphodiesterase* (EC 3.1.4.35) [28]

**Overall druggability result:** Moderate–strong hit. Assigned across **55** *W. bancrofti* sequences.

**Criterion 1, Parasite-specific target:** Global similarity to human homologs is low at **12.65%** sequence identity, comfortably below our selectivity thresholds of <30%.

**Criterion 2, Chemical tractability**: PDEs are established drug targets. We screened a panel of 13 small molecules that spans approved PDE5 inhibitors and research tools: Sildenafil, Tadalafil, Vardenafil, Avanafil, Udenafil, Mirodenafil, Lodenafil, Dipyridamole, Zaprinast, BAY 60-7550, BAY 73-6691, PF-04447943, IBMX (non-selective) [14,15].

**Criterion 3, Biochemical importance:** cGMP signaling regulates muscle tone, sensory–motor behaviors, and fertility in nematodes. A dedicated cGMP-PDE in *W. bancrofti* would imply active cyclic-nucleotide control of movement and reproduction. Inhibiting a parasite PDE would elevate intracellular cGMP and disrupt neuromuscular control, weakening adult roundworms and increasing susceptibility to partner drugs and combination therapy [29].

**Criterion 4, Biologically plausible:** PDE inhibition perturbs nematode physiology and has been proposed as an anti-protozoa concept [30].

Potency in Figure 12 is moderate with sequence-dependent variability; the enzyme is druggable as shown in Figure 13 but will likely need ligand growth & optimization to achieve robust sub-μM activity and nematode-selective profiles.

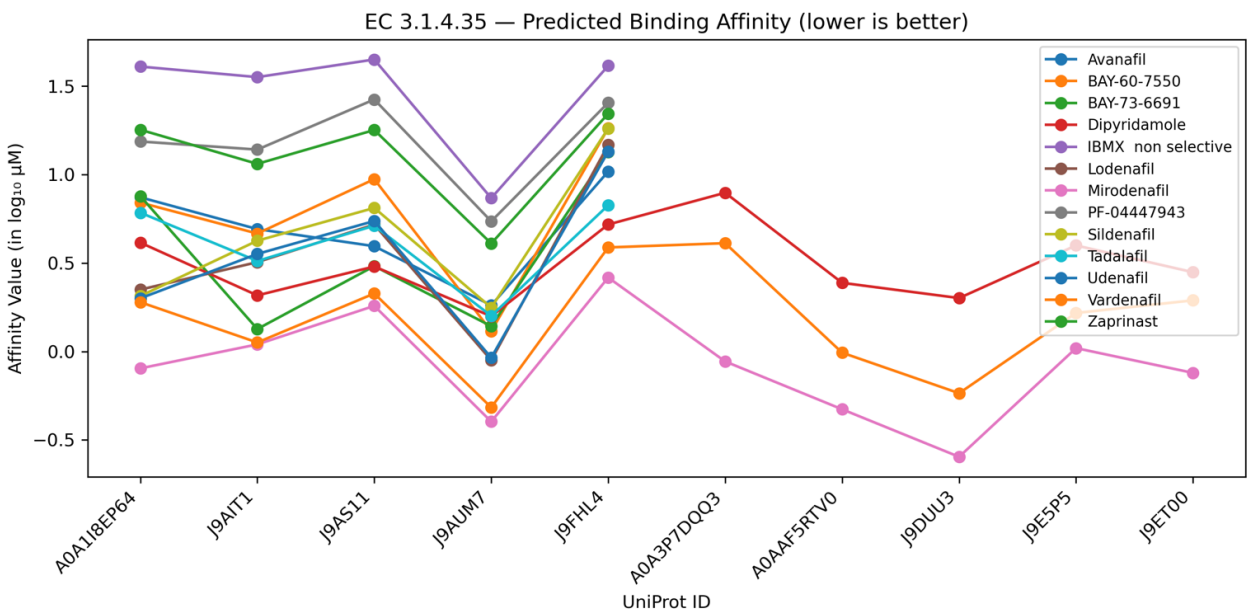

**Figure 12.** Predicted potency of cGMP phosphodiesterase (EC 3.1.4.35) inhibitors

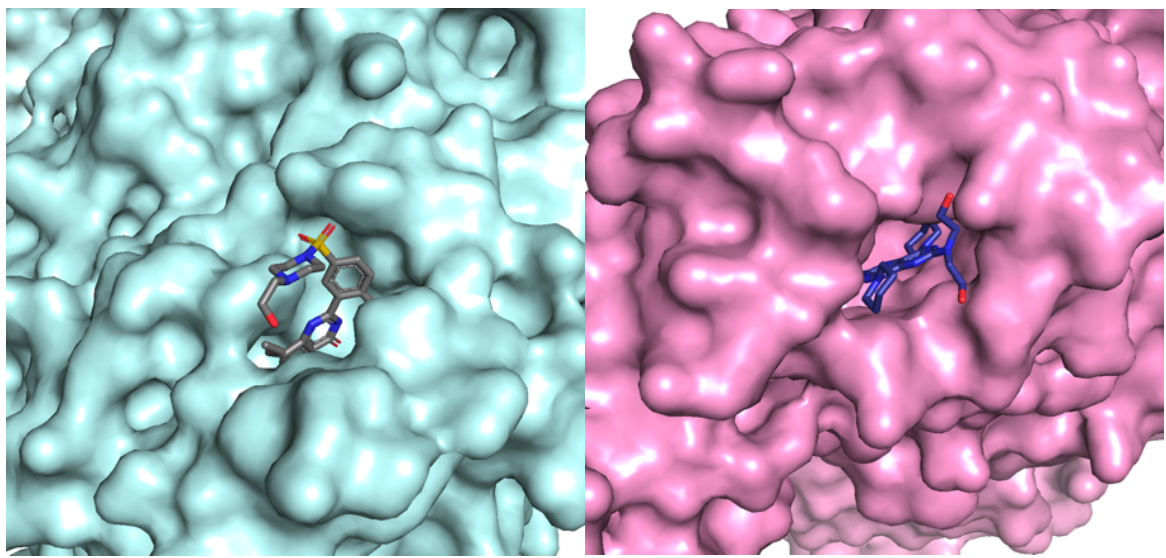

**Figure 13.** Mirodenafil (left) and Dipyridamole (right) complexed with cGMP phosphodiesterase

*3.3 Results Summary*

**Table 2.** Summary of Enzyme Targets and Representative Top Inhibitor Affinities

| Strategy | EC # | Top Inhibitor | IC$_{50}$ Range (nM) | Overall Qualification |
|---|---|---|---|---|
| Anti-Wolbachia | 2.4.99.28 | Moenomycin A12 | 16 – 100 | Strong |
| Core Function | 3.4.22.1/65 | K777 | 244 – 750 | Moderate-Strong |
| Life Cycle | 3.2.1.14 | Allosamidin | 715 – 10000 | Moderate |
| Immune-Evasion | 3.6.1.15 | Suramin | 88 – 500 | Strong |
| Destabilization | 3.1.4.35 | Mirodenafil | 254 – 950 | Moderate-Strong |

Table 2 highlights the top-scoring inhibitor for each target, its approximate binding strength, and overall recommendation using Boltz-2. In summary, the docking study found at least one moderately strong binder for every target. For two targets – the Wolbachia glycosyltransferase and the NTPase – we identified low-nanomolar candidates (moenomycin analogs and suramin, respectively).

## 4. Discussion

*Significance of New Annotations*

Precise EC assignments of the *W. bancrofti* proteome shed light on the underlying biology of this organism. Our pipeline systematically mapped the biochemical roles for thousands of proteins, effectively illuminating metabolic pathways and potential vulnerabilities. Because many of these pathways involve multiple annotated enzymes, they present opportunities for combination therapy to enhance efficacy and reduce drug resistance. These findings lay the groundwork for future experimental studies: researchers can prioritize these newly annotated enzymes for functional assays, vaccine studies, or genetic knockdown in model systems.

*Challenges and Future Directions*

While our pipeline combines many state-of-the-art deep learning methods, further experimental validation is needed to prioritize targets. The next step is to express these target enzymes (or obtain parasite extracts) and perform biochemical inhibition assays with these compounds.

Our approach can also be generalized and improved. For instance, we could apply the same pipeline to other neglected parasites or pathogens with genome sequences but many hypothetical proteins [31]. As a proof of concept, *W. bancrofti* was particularly challenging due to a lack of prior data, so this result is extrapolatable for other organisms. Improvements could include incorporating de novo generative chemistry in the loop. For instance, once we identified suramin as a hit for NTPase, one could use a fragment-based design tool (like DeepFrag [32]) to suggest modifications that improve potency and drug-like properties. We attempted a pilot ligand growing experiment using DeepFrag on some of the docked poses of moenomycin analogs and suramin to see if we could further improve predicted affinity. The initial results were mixed – DeepFrag did propose new fragments that could strengthen interactions. However, often the proposed additions did not maintain the original pose or introduced clashes upon re-docking. This also highlights a need for careful human-in-the-loop when using generative design.

**5. Conclusion**

Our study delivers a prioritized shortlist of anti-filarial targets and a broadly applicable computational framework. By integrating deep learning–based enzyme annotation with structure-based drug discovery, we decoded a previously unannotated "dark proteome" of *W. bancrofti* and linked key enzymes to tractable small-molecule inhibitors. While experimental validation remains essential, our results significantly narrow the search space and accelerate early-stage drug development for *W. bancrofti* and related parasites.

**References**

1. WHO. Lymphatic filariasis [Internet]. Who.int. World Health Organization: WHO; 2024. Available from: https://www.who.int/news-room/fact-sheets/detail/lymphatic-filariasis

2. Rsispostadmin. Lymphatic Filariasis: Insightful Review of a Neglected Tropical Disease - RSIS International [Internet]. RSIS International. 2023 [cited 2025 Aug 7]. Available from: https://rsisinternational.org/virtual-library/papers/lymphatic-filariasis-insightful-review-of-a-neglected-tropical-disease

3. Thomsen EK, Sanuku N, Baea M, Satofan S, Maki E, Lombore B, et al. Efficacy, Safety, and Pharmacokinetics of Coadministered Diethylcarbamazine, Albendazole, and Ivermectin for Treatment of Bancroftian Filariasis. Clinical Infectious Diseases. 2015 Oct 20;62(3):334–41.

4. Zhu C, Yan Y, Feng Y, Sun J, Mu M, Yang Z. Genome-Wide Analysis Reveals Key Genes and MicroRNAs Related to Pathogenic Mechanism in Wuchereria bancrofti. Pathogens [Internet]. 2024 Dec 10 [cited 2025 Jan 26];13(12):1088. Available from: https://www.mdpi.com/2076-0817/13/12/1088

5. UniProt [Internet]. Uniprot. 2023. Available from: https://www.uniprot.org/

6. ExPASy - ENZYME [Internet]. enzyme.expasy.org. Available from: https://enzyme.expasy.org/

7. Ryu JY, Kim HU, Lee SY. Deep learning enables high-quality and high-throughput prediction of enzyme commission numbers. Proceedings of the National Academy of Sciences. 2019 Jun 20;116(28):13996–4001.

8. Gi Bae Kim, Ji Yeon Kim, Jong An Lee, Norsigian CJ, Palsson BO, Sang Yup Lee. Functional annotation of enzyme-encoding genes using deep learning with transformer layers. Nature Communications. 2023 Nov 14;14(1).

9. Trentini B, Lorenz P. *Breakthrough in functional annotation with HiFi-NN.* NVIDIA Technical Blog. (2023). Available from: https://developer.nvidia.com/blog/breakthrough-in-functional-annotation-with-hifi-nn/

10. Passaro S, Corso G, Wohlwend J, et al. *Boltz-2: Towards accurate and efficient binding affinity prediction.* (MIT Jameel Clinic Technical Report, 2025). Available from: https://boltz.bio/boltz2

11. Yang Z, Su B, Chen J, Wen JR. Interpretable Enzyme Function Prediction via Residue-Level Detection [Internet]. arXiv.org. 2025. Available from: https://arxiv.org/abs/2501.05644

12. Weights and Biases, Inc [Internet]. Weights & Biases. 2024. Available from: https://wandb.ai/site/

13. NCBI. BLAST: Basic Local Alignment Search Tool [Internet]. Nih.gov. 2025. Available from: https://blast.ncbi.nlm.nih.gov/Blast.cgi

14. ChEMBL Database [Internet]. www.ebi.ac.uk. Available from: https://www.ebi.ac.uk/chembl/

15. Binding Database Home [Internet]. www.bindingdb.org. Available from: https://www.bindingdb.org/rwd/bind/index.jsp

16. Landmann F, Voronin D, Sullivan W, Taylor MJ. Anti-filarial Activity of Antibiotic Therapy Is Due to Extensive Apoptosis after Wolbachia Depletion from Filarial Nematodes. Schneider DS, editor. PLoS Pathogens. 2011 Nov 3;7(11):e1002351.

17. ENZYME - 2.4.99.28 peptidoglycan glycosyltransferase [Internet]. Expasy.org. 2025 [cited 2025 Aug 7]. Available from: https://enzyme.expasy.org/EC/2.4.99.28

18. Nygaard R, Chris, Meagan Belcher Dufrisne, Colburn JD, Pepe J, Hydorn MA, et al. Structural basis of peptidoglycan synthesis by E. coli RodA-PBP2 complex. Nature communications. 2023 Aug 24;14(1).

19. ENZYME - 3.4.22.1 cathepsin B [Internet]. Expasy.org. 2025 [cited 2025 Aug 7]. Available from: https://enzyme.expasy.org/EC/3.4.22.1

20. ENZYME - 3.4.22.65 peptidase 1 (mite) [Internet]. Expasy.org. 2025 [cited 2025 Aug 7]. Available from: https://enzyme.expasy.org/EC/3.4.22.65

21. Ford L, Zhang J, Liu J, Hashmi S, Fuhrman JA, Oksov Y, et al. Functional Analysis of the Cathepsin-Like Cysteine Protease Genes in Adult Brugia malayi Using RNA Interference. PLOS Neglected Tropical Diseases. 2009 Feb 10;3(2):e377–7.

22. ENZYME - 3.2.1.14 chitinase [Internet]. Expasy.org. 2024. Available from: https://enzyme.expasy.org/EC/3.2.1.14

23. Wu Y, Preston G, Bianco AE. Chitinase is stored and secreted from the inner body of microfilariae and has a role in exsheathment in the parasitic nematode Brugia malayi. Molecular and Biochemical Parasitology. 2008 Sep;161(1):55–62.

24. Quek S, Cook DAN, Wu Y, Marriott AE, Steven A, Johnston KL, et al. Wolbachia depletion blocks transmission of lymphatic filariasis by preventing chitinase-dependent parasite exsheathment. Proceedings of the National Academy of Sciences. 2022 Apr 4;119(15).

25. ENZYME - 3.6.1.15 nucleoside-triphosphate phosphatase [Internet]. Expasy.org. 2025 [cited 2025 Aug 7]. Available from: https://enzyme.expasy.org/EC/3.6.1.15

26. Maizels RM, Smits HH, McSorley HJ. Modulation of Host Immunity by Helminths: The Expanding Repertoire of Parasite Effector Molecules. Immunity. 2018 Nov;49(5):801–18

27. Lisvane Paes-Vieira, André Luiz Gomes-Vieira, José Roberto Meyer-Fernandes. E-NTPDases: Possible Roles on Host-Parasite Interactions and Therapeutic Opportunities. Frontiers in Cellular and Infection Microbiology. 2021 Nov 9;11.

28. ENZYME - 3.1.4.35 3',5'-cyclic-GMP phosphodiesterase [Internet]. Expasy.org. 2025 [cited 2025 Aug 7]. Available from: https://enzyme.expasy.org/EC/3.1.4.35

29. Schuster KD, Mohammadi M, Cahill KB, Matte SL, Maillet AD, Vashisth H, et al. Pharmacological and molecular dynamics analyses of differences in inhibitor binding to human and nematode PDE4: Implications for management of parasitic nematodes. Koch KW, editor. PLOS ONE. 2019 Mar 27;14(3):e0214554.

30. Maurice DH, Ke H, Ahmad F, Wang Y, Chung J, Manganiello VC. Advances in targeting cyclic nucleotide phosphodiesterases. Nature reviews Drug discovery [Internet]. 2014 Apr 1;13(4):290–314. Available from: https://www.ncbi.nlm.nih.gov/pmc/articles/PMC4155750/

31. Sathyanarayanan N, Nagendra H. Genome wide survey and molecular modeling of hypothetical proteins containing 2Fe-2S and FMN binding domains suggests Rieske Dioxygenase Activity highlighting their potential roles in bio-remediation. Bioinformation [Internet]. 2014 Feb 19 [cited 2025 Aug 7];10(2):68–75. Available from: https://pmc.ncbi.nlm.nih.gov/articles/PMC3937578/

32. Green H, Durrant JD. DeepFrag: An Open-Source Browser App for Deep-Learning Lead Optimization. Journal of Chemical Information and Modeling. 2021 May 24;61(6):2523–9